\documentclass[twocolumn]{aastex631}

\shorttitle{Variability of  the highest redshift blazar}
\shortauthors{Moretti et al.}
\graphicspath{{./}}

\begin{document}

\title{Minute-timescale variability in the X-ray emission of  the highest redshift blazar.}

\correspondingauthor{Alberto Moretti}
\email{alberto.moretti@inaf.it}

\author[0000-0002-9770-0315]{Alberto Moretti}
\affiliation{ INAF, Osservatorio Astronomico di Brera \\
Via Brera 28  \\
20121 Milano, Italy}

\author[0000-0002-0037-1974]{Gabriele Ghisellini}
\affiliation{ INAF, Osservatorio Astronomico di Brera 
Via  E. Bianchi 46 
23807 Merate (LC), Italy}

\author[0000-0002-2339-8264]{Alessandro Caccianiga}
\affiliation{ INAF, Osservatorio Astronomico di Brera \\
Via Brera 28  \\
20121 Milano, Italy}

\author[0000-0003-4747-4484 ]{Silvia Belladitta}
\affiliation{
DiSAT, Universit\`a degli Studi dell'Insubria
Via Valleggio 11
22100 Como, Italy}
\affiliation{ INAF, Osservatorio Astronomico di Brera \\
Via Brera 28  \\
20121 Milano, Italy}

\author[0000-0001-7551-2252]{Roberto Della Ceca}
\affiliation{ INAF, Osservatorio Astronomico di Brera \\
Via Brera 28  \\
20121 Milano, Italy}

\author[0000-0003-1516-9450]{Luca  Ighina} 
\affiliation{
DiSAT, Universit\`a degli Studi dell'Insubria
Via Valleggio 11
22100 Como, Italy}
\affiliation{ INAF, Osservatorio Astronomico di Brera \\
Via Brera 28  \\
20121 Milano, Italy}

\author[0000-0002-3069-9399]{ Tullia  Sbarrato} 
\affiliation{ INAF, Osservatorio Astronomico di Brera 
Via  E. Bianchi 46 
23807 Merate (LC), Italy}

\author[0000-0001-5619-5896]{ Paola  Severgnini} 
\affiliation{ INAF, Osservatorio Astronomico di Brera \\
Via Brera 28  \\
20121 Milano, Italy}

\author[0000-0002-2231-6861]{Cristiana  Spingola} 
\affiliation{ INAF, Istituto di Radioastronomia 
Via Gobetti 101
40129 Bologna, Italy }

\begin{abstract}
We report on two Chandra observations  of the quasar PSO J0309+27, the  most distant blazar observed so far (z=6.1),
performed eight months apart, in March and November 2020.  Previous Swift-XRT observation showed that this object  is one of the  brightest X-ray 
sources beyond redshift 6.0 ever observed so far.  This new data-set confirmed the high flux level and unveiled a spectral change  
occurred on a very short timescale (250s rest-frame), caused by a significant softening of the emission spectrum.  
This  kind of spectral variability, on a such short interval, has never been reported in the X-ray emission of a flat spectrum radio quasar. 
A possible  explanation is given by the emission produced by the inverse Compton scatter of the quasar UV photons by 
the cold electrons present in a fast shell moving along the jet.  Although this bulk comptonization emission should be an unavoidable consequence 
of the standard leptonic jet model, this would be the first time that it is observed. 
\end{abstract}

\section{Introduction} \label{sec:intro}
     
Radio loud quasars (RL QSOs)  are characterised by the presence of  relativistic jets, huge collimated outflows originating very close to
the accreting supermassive black hole (SMBH) and extending up to extragalactic distances \citep{Harris06, Blandford20}. 
When aligned with our line of sight, the jet overwhelms most of the emission from the other nuclear components.
In this case, RL QSOs are called blazars and their typical spectral energy distribution (SED)  shows two main peaks, the 
first falling between the IR and the X-ray energy bands and the second at the 
gamma-ray frequencies \citep[see] [for a recent review]{Madejski16}. According to the standard picture, the low energy component 
 is synchrotron  radiation  produced by a population of electrons moving  at highly  relativistic velocities in random directions within the jet flow.
The high energy photons are thought to be  produced by the so called Self Synchrotron Compton (SSC) mechanism, which is  
the inverse Compton (IC) scatter of the same electrons on the low energy photons produced via synchrotron mechanism,  \citep[e.g.]{Sikora97, Ghisellini09}.  
The flat spectrum radio quasar (FSRQ), which are the blazars hosted by QSOs,  are characterised by dense radiative circumnuclear environment. 
In these cases other possible inverse Compton (IC) seed photons can be external to the jet, like the ones produced in the accretion disk and reprocessed by the
 broad line regions (BLR) or by the dusty torus. At larger scale  stellar radiation of the host galaxy and the Cosmic Microwave Background  
 (CMB) photons on extragalactic scale are expected to contribute \citep{Tavecchio00, Harris06}.
\begin{figure}
\center{
\includegraphics[width=6.5cm] {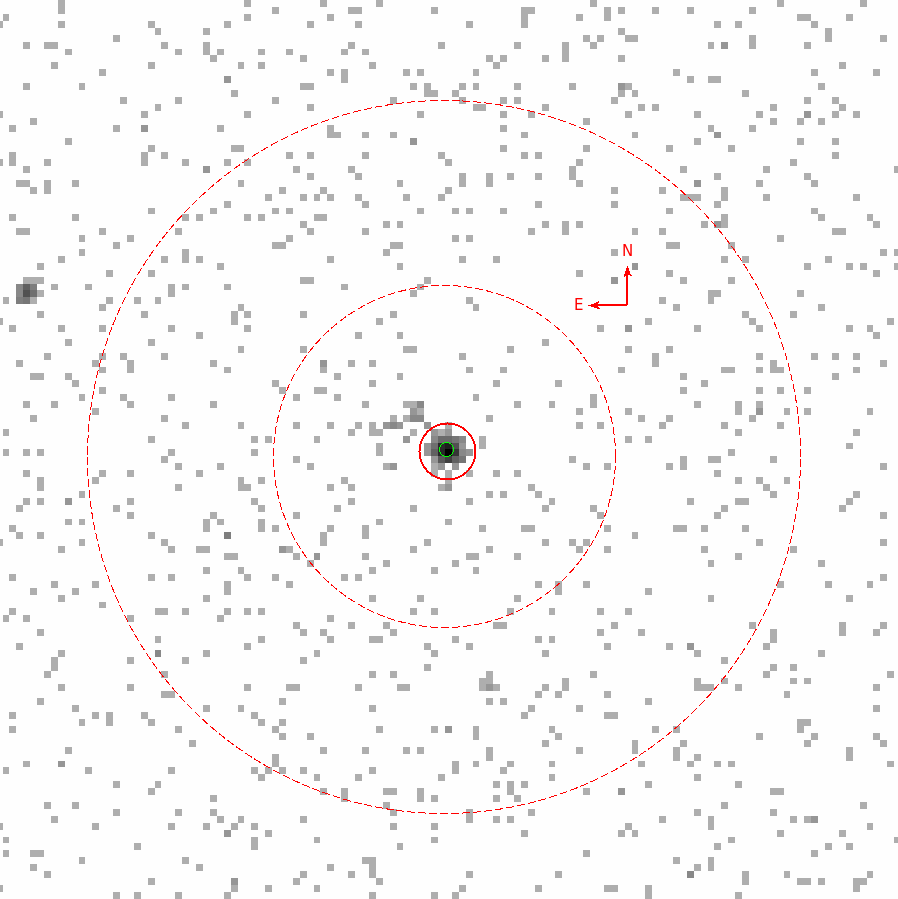} 
\caption{Chandra image of PSO J0309+27. Red small circle and red-dashed annuli show the source and background extraction regions
(2'', 12'', 25"  radii respectively)  used for temporal and spectral analysis. Green circle draws the position of the PS1 optical counterpart.
 An extended emission is clearly visible on the NE direction from the
core: its properties will be presented in a forthcoming paper.}
}
\label{fig:imaz}
\end{figure}

While multi-wavelength observations generated a broad consensus in the literature on the main emission mechanisms, 
there are some fundamental aspects of the jet nature  that still escape understanding.
One of the most basic issue is their composition: a large amount of energy is observed moving through the jet from the very central part 
of the QSO up to Megaparsec scale. Energy can be  transported by a combination of leptons, protons and 
Poynting flux. It is not clear what is the relative contribution of the different elements and how it depends on
the distance from the SMBH \citep{Sikora05}.
Determining the composition would allow us to calculate the jet total power and to assess its impact on the QSO environment. 
Moreover knowing the baryon and lepton loads of the jets would be useful to ascertain their formation mechanism \citep{Sikora00}.
The investigation of the spectral variability is a very effective  tools to probe the properties and the dynamics  of the radiating particles,
and to gain some insight on  the  physical processes responsible for the observed emission.  
\begin{table}
\center{
\caption{(i) Epoch of the observations; (ii) Chandra archive ID; (iii) starting date; (iv) duration. 
Note that the November observation has been splitted in 5 different segments. \\}
\begin{tabular}{c|c|c|c}
\hline
\hline
Epoch&Obs Id.&Date & Exp[s]   \\
\hline
 March&   23107   &    2020-03-24T20:35:21    &    26,703 \\
\hline
 November &    24513    &    2020-11-03T10:02:02    &     19,643  \\
&    24855    &    2020-11-04T00:31:18    &    8,648 \\
&    23830    &     2020-11-05T14:30:34   &    21,794 \\
&    24856    &    2020-11-06T11:46:48   &    23,414  \\
&    24512    &    2020-11-07T14:30:30   &     27,713  \\
\hline
\end{tabular}
\label{tab:obs} 
}
\end{table}

In this paper we report on the Chandra observation of PSO J030947.49+271757.31, hereafter PSO J0309+27, 
the most powerful radio-loud QSO at redshift higher than 6  (z=6.1).
The original aim of the Chandra observation was to ascertain the presence and the characteristics of the jet emission.
The imaging data analysis together with the properties of the kpc scale  X-ray emission will be presented in a forthcoming paper (Ighina et al., in prep.).
Here we focus on the core emission in which we detected a significant spectral variation on the surprisingly short observed  time scale of $\sim$ 2 ks. 
While other convincing explanations are lacking in the literature, these data are close to what had been predicted 
in a scenario of bulk comptonization (BC).  
In Sect.~\ref{sect:obj} we briefly describe the multi-wavelength database currently available for the this source.  
In Sect.~\ref{sect:data} we present the Chandra data timing and spectral analysis.
In Sect.~\ref{sect:disc} we discuss the interpretation of this observation as due to the BC  of  UV photons of the QSO by the 
cold electrons of the jet.
    
Throughout this paper, errors are quoted at 68\% confidence level, unless otherwise specified.
We adopt the following cosmology parameter values:
$\Omega_{\rm m} = 0.3$, $\Omega_\Lambda = 0.7$, $h_0 = 0.7$.
\begin{figure*}
\includegraphics[width=9.3cm]   {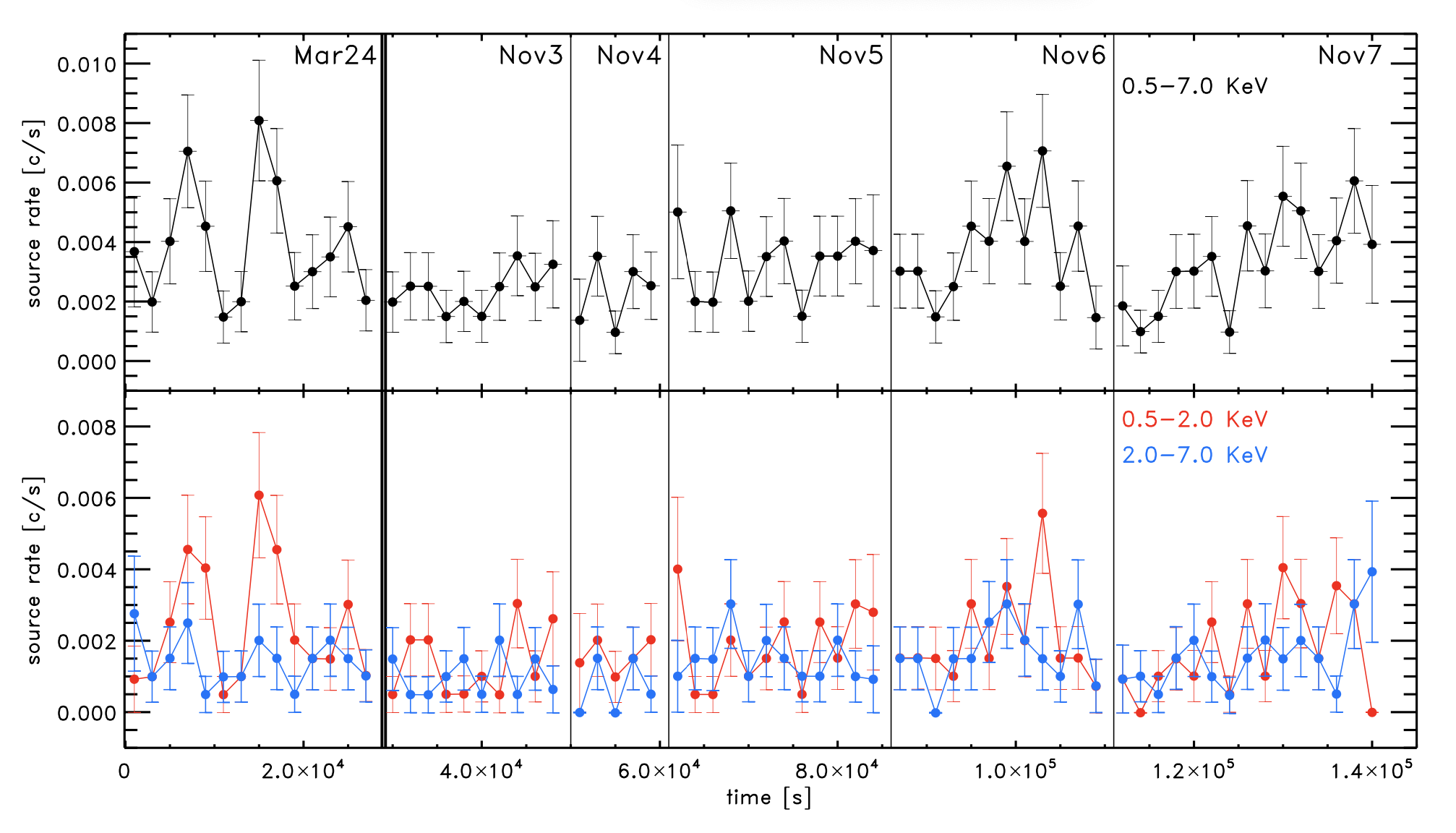}   \includegraphics[width=8.7cm] {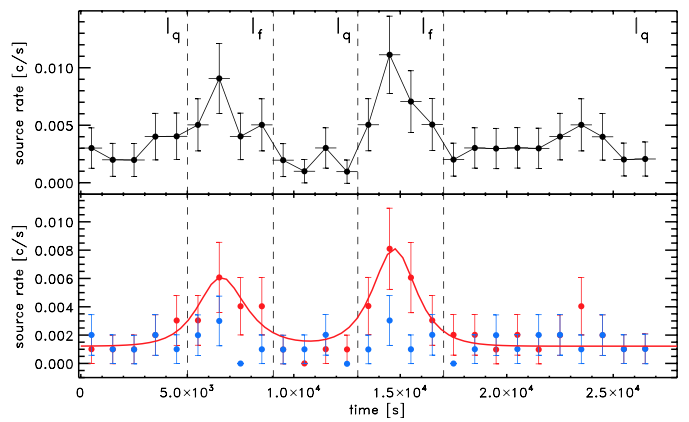} 
\caption{ {\bf Left Panel:} Background subtracted  light curve of the whole observation with 2 ks bin. The bars are  Poisson error
as calculated by the \texttt{dmextract} task.  For the sake of clarity, 
bin times are represented in a continuous way, although data have been collected in 6 different and unconnected intervals. 
First observation is kept apart by a double continuous line, while the segments of the second observation are 
divided by dashed lines. Data-sets are labelled in accordance with Tab.\ref{tab:obs}. 
The total band data are shown in the upper panel, hard and soft bands in the lower panel in blue and red respectively. 
 {\bf Right Panel:} light curve of the first observation only in temporal bins of 1 ks. Continuous red line shows the fit to the 
 flares according the model described in the text. Dashed vertical black  lines show the flaring intervals used for the time 
 resolved spectroscopy.}
\label{fig:lc}
\end{figure*}
%

\section{PSO J0309+27}\label{sect:obj}

PSO J0309+27 was selected by combining the NRAO VLA Sky Survey, NVSS \citep{Condon98}  and the  Panoramic
Pan-STARRS \cite[PS,] []{Chambers16}  catalogs  and spectroscopically confirmed  as a z=6.1 QSO using the Large Binocular Telescope (LBT) 
in October 2019 \citep{Belladitta20}.  At z$>$6 this is by far the most luminous QSO in the radio band.

A relatively short (19 ks) Swift observation performed between October and November 2019, detected the source with 
a  flux of $\sim$ 3.4 $\times$ 10$^{-14}$ erg s$^{-1}$ cm$^{-2}$  in the 0.5-10.0 energy band, corresponding to a luminosity 
of  $\sim$10$^{45}$ erg s$^{-1}$  in the 2-10 keV band. 
This allowed us to determine that the X-ray emission is well above the expected coronal emission and it is, with
 high probability, due to a jet pointing in our direction \citep{Belladitta20}.  

PSO J0309+27 has been observed  at the 1.5, 5 and 8.4 GHz frequencies with VLBA in April 2020
under a DDT project \citep{Spingola20}. The milliarcsecond angular resolution revealed 
a 500 pc one-sided bright jet resolved into several components.
In combination with the core dominance value \citep{Giovannini94}  these data 
constrain the jet Lorentz factor  between  3-5 at small viewing angles ($\rm \theta_{view} < 10^{\circ}$).

PSO J309+27 has been also observed  in the near-IR with the LBT-LUCI spectrograph. 
From this observation Belladitta et al. (2021, in preparation) measures a mass of  8$^{+3}_{-2}\times$10$^8 M_{\odot}$
corresponding  to a gravitational radius of  $\sim$1.2 $\times$10$^{14}$ cm and 
a  135 nm luminosity  of  1.1$\pm$0.2$\times$ 10$^{46}$ erg s$^{-1}$,
corresponding to a  BLR radius of  $\sim$ 2 $\times$ 10$^{17}$ cm \citep{Lira18}. 
The estimated Eddington ratio is $\sim$ 0.3 .

\section{Observation and data reduction}\label{sect:data}

Chandra first observation of PSO J0309+27  was performed under Directory Discretionary Time  (DDT 704032) on March 24 2020 for a total 
effective time of 26.7 ks (	Archive Seq. Num. 704032, PI A.Moretti).
Then, a  100 ks observation was approved as part of the Cycle 22. Data were collected in 5 different exposures 
between Nov. 3rd and Nov. 7th 2020  (Tab. \ref{tab:obs}) for a total effective time of 101.2 ks (Archive Seq. Num. 704242, PI A.Moretti).   
Both the observations were conducted with ACIS-S3 detector (CCD=7) set to very faint mode, in order to reduce the background.  
No data loss  occured due to soft protons flares. We reprocessed the data by using \texttt{chandra\_repro} script in CIAO 4.12.1 
\citep{Fruscione06} and using  CALDB 4.9.3 calibration library.   
The source is detected with 418 photons included in a 1.5" radius circle in the 0.5-7.0 keV energy band , with 1.9 background events expected. 
This makes the background negligible for the purposes of the following analysis.

\subsection{Spectral analysis of the two observations} \label{sect:spec0}
    
In both the March and November observations the source spectrum was extracted by means of the \texttt{specextract} CIAO task, 
using a 2" radius circular region centered on the position as measured  by the \texttt{wavedetect} tool. Auxiliary response (ARF) and 
response matrix (RMF) files  have been produced by the same task. The background was estimated in an annulus centered in the 
same position with 12" and 25" as internal and external radii (Fig. \ref{fig:imaz}).

Data  were fitted in XSPEC (version 12.10.1)  using the C-statistics with a simple absorbed power law with the absorption factor fixed to the 
Galactic value (1.16$\times$10$^{21}$ cm$^{-2}$) as measured by the HI Galaxy map (Kalberla et al. 2005).
 The photon index shows a significant  variation from $\Gamma$=2.10$\pm$ 0.20 in the first observation (March) to $\Gamma$=1.63$\pm$ 0.10 in the second (November).
This latter value is in agreement with previous Swift observation \citep{Belladitta20} and consistent with the typical FSRQ values, for which an X-ray photon index of  $\lesssim$ 1.8 is expected \citep{Ighina19}; 
on the contrary the spectrum observed in  March is significantly softer. The observed flux in the 0.5-10.0 band did not significantly change between the two epochs (Tab.\ref{tab:spec_fit}). 
    
To measure the goodness of the fit  we followed the approach of \cite{Medvedev20}, using the Anderson-Darling (AD) test statistic (XSPEC manual)
\footnote{https://heasarc.gsfc.nasa.gov/xanadu/xspec/manual}. By means of the  XSPEC \texttt{goodness} task (with "nosim" and "fit"  options) 
we re-sampled both the March and November best fit models 10,000 times.  
We found that  74\%  and 92\% of the simulated data sets have a better test statistics, respectively. Since the  goodness-of-fit tests only allows us to reject a model,
this means that  power law is consistent with both data-sets.     

\subsection{Temporal  analysis} 
     
To better investigate the spectral variation between the 2 observations,  using \texttt{dmextract} task, we produced the light curve in 
temporal bin of 2000 seconds, in three different energy bands:  wide (0.5-7.0 keV),  soft (0.5-2.0 keV) and hard (2.0-7.0 keV).  
The source  and background data were extracted from the same extraction regions used in the  spectral analysis.
As shown in Fig.~\ref{fig:lc}, the background subtracted wide-band light curve, during the March observation, two intervals of  approximately 
4,000 s each (between 5-9 ks and 13-17 ks from the observation start) present a flux level which is higher with respect to the rest of the observation. 
As it is clear from the comparison between soft and hard band curves this "flaring activity " is almost entirely restricted to energy below 2.0 keV.

To give a first non-parametric estimate of the statistical significance of the observed variation we made use of a 2-dimensional Kolmogorv-Smirnoff test 
\citep[KS2D][and references therein]{Press02}.  We compared the photon observed inter-arrival time and energy distributions with synthetic samples generated from 
non variable models. For the inter-arrival times $\Delta$t we used the  exponential distribution e$\rm^{-cr \Delta t }$, where cr is the mean observed count-rate,
as expected in a Poisson process\citep{Feigelson13}.  For the energy distribution we used the simple power-law model giving the best fit to the whole data-set 
\ref{tab:spec_fit}. We found that  2\% of 10,000  synthetic data-set exceed the distance of our observation from the model. This gives to the observed 
variability a non parametric statistical significance of 98\%.

In order to give a rough estimate of the characteristic  observed variability timescale we qualitatively    
modelled  the 1 ks binned soft light-curve with  the following simple analytical function \citep[][and reference therein]{Hayashida15}
\begin{equation}
\rm F(t) \propto \frac {1} {e^{-(t-t_0)/\tau_{rise}} + e^{-(t-t_0)/\tau_{fall}}}.
\end{equation}
We found that the rising and falling parts ($\rm \tau_{raise}$ and $\rm \tau_{fall}$) are $\sim$ 900 s, 
as shown in the right-bottom panel of  Fig.~\ref{fig:lc}.   

In the following we focus on the time resolved spectral analysis of the first observation:
as we will see, the difference in photon index between the two epoch is entirely restricted to these two episodes.  

\subsection{Time-resolved spectral analysis}
\begin{figure}
\center{
\includegraphics[width=9cm] {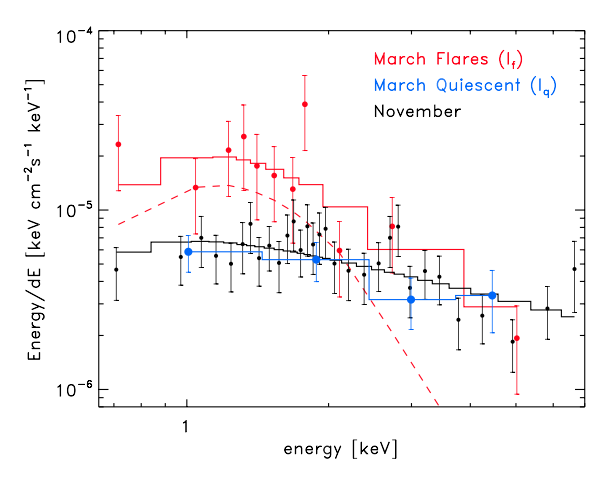}      
\caption{ The unfolded spectrum extracted in the  I$\rm_f$,  I$\rm_q$  intervals of the March  and in the November observation. 
Spectral data are plotted as points with vertical error bars. Best fit models are plotted with continuous lines. 
Red and blue points show flaring  (I$\rm_f$) and quiet (I$\rm_q$) intervals respectively. The latter have been modelled by a 
simple power-law (continuous line). 
I$\rm_f$ have been fitted by the sum of the  I$\rm_q$ power-law plus a thermal component shown by the red dashed line. 
November data, in black,  are consistent with  I$\rm_q$ data. 
Data are binned at 2$\sigma$ significance only for graphical purposes.}
\label{fig:spec}
}
\end{figure}

In order to characterise the spectral changes, first, we analysed the March observation data accumulated during the flares and during the
quiescent periods separately.  We considered as flaring the intervals 5095-9127s and 13095-17127s from the beginning of the observation (which took place at 
701470232.65 satellite time), corresponding to an exposure time of 7959s.  
For the seek of simplicity hereafter we will refer to the quiescent and flaring intervals as I$\rm_q$ and I$\rm_f$ respectively
(Fig. \ref{fig:lc}).
Using the two-sample KS test \citep{Press02}, we found that the photon energy distribution observed in the  I$\rm_f$ interval is different from the rest of the observation, with 
a confidence of  99.35\% (KS probability =  0.0065).

\begin{table*}
\caption{Results of the spectral analysis. Single power law best fit values of the following data-sets are reported:
March observation (26.7 ks); November observation (101.7 ks); March quiescent intervals (I$\rm_q$ , 18.7 ks); March flare intervals (I$\rm_f$, 7.9 ks). 
For each data-set we report: (ii) the photon index;  two band  fluxes (iii-iv) and luminosities (v-vi), corrected accounting for the Galactic absorption. \\} 
\center{
\begin{tabular}{ccccccc}
\hline
\hline
                 
Epoch        &   ph. index                              & flux$\rm_{0.5-2 keV}$ & flux$\rm_{2-10 keV}$  & lum$\rm_{2-10 keV}$  & lum$\rm_{15-50 keV}$  & dof/cstat (goodness)    \\         
             &                           & 10$^{-14}$erg s$^{-1}$cm$^{-2}$ & 10$^{-14}$erg s$^{-1}$cm$^{-2}$ & 10$^{45}$erg s$^{-1}$ & 10$^{45}$erg s$^{-1}$ & \\
(i) &(ii) &(iii) &(iv) &(v) &(vi) &(vii) \\ 
\hline 
Mar          &  2.11$_{-0.20}^{+0.20}$    & 3.23$_{-0.52}^{+0.47}$  & 3.18$_{-0.51}^{+0.46}$  & 16.78$_{-2.73}^{+2.43} $ &  14.65$_{-2.38}^{+2.12} $  & 77/67.21   (74\%) \\                    
Nov          &  1.66$_{-0.11}^{+0.11}$    & 1.98$_{-0.20}^{+0.186}$ & 3.86$_{-0.39}^{+0.36}$  &  8.46$_{-0.85}^{+0.79} $ &  15.10$_{-1.53}^{+1.42} $  & 176/128.81 (92\%) \\                  
I$\rm_q$     &  1.71$_{-0.26}^{+0.27}$    & 1.83$_{-0.46}^{+0.39}$  & 3.30$_{-0.83}^{+0.70}$  &  7.99$_{-2.00}^{+1.71} $ &   3.25$_{-3.32}^{+2.82}$   & 44/37.62   (54\%) \\                  
I$\rm_f$     &  2.57$_{-0.30}^{+0.30}$    & 6.97$_{-0.15}^{+0.13}$  & 3.47$_{-0.75}^{+0.65}$  & 44.34$_{-9.63}^{+8.36} $ &  18.78$_{-4.08}^{+3.54}$   & 43/41.98    (47\%) \\                    
\hline
\end{tabular}
}
\label{tab:spec_fit} 
\end{table*}
Both I$\rm_q$  and  I$\rm_f$ data can be well fitted by single power law models.
The best fit slope value for I$\rm_q$ is  1.70$_{-0.26}^{+0.27}$, closer to and consistent with the second epoch (November) values. 
I$\rm_f$ data are significantly softer with a  photon index best value  of 2.58$_{-0.30}^{+0.30}$.
Using the method described in Sect.\ref{sect:spec0} we reject the possibility that the two datasets are consistent with a confidence level $>$99.99\%.
Indeed, assessing the goodness of the fit by the same single power law,  we found that 100\% out of the 10,000 simulated data-sets have better statistics.
Source flux  in the 0.5-10 keV band doubles during the flaring phase, going  from 5.01$_{-0.66}^{+0.93} \times$10$^{-14}$ erg s$^{-1}$ cm$^{-2}$
 during  I$\rm_q$ up to 1.05$_{-0.14}^{+0.16} \times$10$^{-13}$ erg s$^{-1}$ cm$^{-2}$ during the I$\rm_f$, while the luminosity in the 2.0-10. keV 
 band (0.3-1.4 rest-frame) increased by a factor 5.5.
\begin{table*}
\caption{Result of the spectral fit to the transient component, when  I$\rm_f$ data are fitted by 
the sum of a BB and power law with parameters frozen to the I$\rm_q$ values. We list:
(i) the BB mean temperature, (ii) flux and   (iv) luminosities. Flux and luminosity have been corrected accounting for the Galactic absorption. } 
\center{   
\begin{tabular}{cccc}
\hline
\hline
 KT         &   flux$\rm_{0.5-2keV}$                     &     lum$\rm_{2-10keV}$  &    C-stat/dof (good.)  \\
KeV        &  10$^{-14}$erg s$^{-1}$cm$^{-2}$  &  10$^{45}$erg s$^{-1}$   &                                  \\
     (i)      &     (ii)                                                 &    (iii)                               &   (iv)                           \\
\hline	  
   2.39$_{-0.45}^{+0.53} $    &      3.32$_{-0.76}^{+0.93}$  &     11.67$_{-2.75}^{+3.14} $&    41.53/43 (17\%)   \\
\hline 
\end{tabular}
}
\label{tab:spec_bb} 
\end{table*}
Rather than with a change in the total power-law slope, the same data can be interpreted as due to a transient soft emission. 
Aiming at testing the  BC emission hypothesis, we assumed that the I$\rm_f$ is the sum of 
the quiescent component, observed in  I$\rm_q$, plus  an extra component, which we modelled  as a Black Body (BB) as expected in 
such a scenario \citep{Celotti07}.  
We fit the I$\rm_f$ data as the sum of a power-law and a BB, freezing the power-law parameters to the to the I$\rm_q$ best fit values (Fig.~\ref{fig:spec}). 
The BB model KT best fit value is 0.33$_{-0.06}^{+0.07}$ keV, corresponding to a rest-frame value of 2.34 $_{-0.43}^{+0.50}$ keV,
consistently with what expected from BC scattering of cold jet electrons onto Ly$\alpha$ photons.  
Flux and luminosity measures are reported in   Tab.\ref{tab:spec_bb}.  

We estimated the goodenss-of-the-fit following the same procedure  described in the previous section for both models, the single power law and 
the frozen power-law plus a black body. The results of the bootstrapping are reported in 
the respective tables: we conclude that neither of the two hypotheses can be discarded on a statistical basis from our data.
   
We note that  the BB parameter estimate is not strongly dependent on definition of the flaring/quiescent intervals.
Indeed, the  spectral slope measured during the March quiescent intervals I$\rm_q$ is fully consistent with the November data 
(Tab.\ref{tab:spec_fit}). Freezing the power law parameters to the best fit of  the I$\rm_q$ March intervals or to the November observation 
does not significantly affect the BB spectral analysis results.  We also note that a soft flare, with properties similar to the 
two episodes registered in March, but with slightly lower  statistical significance, is possibly present at $\sim$ 17 ks from the start of the Nov 6th segment.  
Including it in the I$\rm_f$ would not significantly modify neither the results of  the time resolved spectral analysis, nor their interpretation and the following discussion.
%
\section{Discussion}\label{sect:disc} 

As reported in the previous Section the soft X-ray emission of PSO J0309+27 has been observed varying in two short time
intervals following a softer-when-brighter pattern on  timescale $\rm \Delta \tau_{obs} \simeq \rm \tau_{raise} +\tau_{fall}  $ = 
1800s  corresponding to $\sim$ 250s rest-frame. 

Although tens of years of studies showed that blazars exhibit complex spectral variability on  vast range of time-scales,
to our knowledge, a change so rapid and with these characteristics has never been described in the X-ray  emission of a FSRQ. 
The intra-day variability reported by \cite{Bhatta18} for 7 FSRQs is smaller in amplitude and different 
in spectral behaviour, being harder when brighter. 
Fermi LAT observations of several FSRQ put in evidence  $\gamma$-ray flux variations on few hundreds seconds time scale 
\citep{Ackermann16, Shukla20}; however the analysis of the long-term observational campaigns in the X-ray bands has 
never revealed a similar behaviour \citep[e.g.][]{Hayashida15, Larionov20}.

A variable BB emission is  what  is expected in a BC scenario, where the jet cold electrons up-scatter external 
UV photons to X-ray energies.  This mechanism has been proposed as an observational probe of the lepton content of the blazar jets  
by \cite{Begelman87} and subsequently deeply investigated by several different authors in the literature 
\citep[e.g.][]{Sikora97, Moderski04, Georganopoulos05, Celotti07}. 
Together  with  the highly relativistic lepton population, responsible for the SSC emission,  
the jet should be populated  by a commensurate number of electrons which are not relativistic in the co-moving frame. 
Streaming through an external radiation field  with a Lorentz factor  $\sim ~ \Gamma_{\rm bulk} $,  
these cold electrons would up-scatter photons, producing the so 
called BC emission. In the SMBH proximity, where jet is thought to be launched,  the main external  photon source 
is the accretion disk. In particular the main contribution to BC emission is expected by those disk photons  
re-emitted by BLR toward the SMBH,  since they are seen head-on  in the 
 jet frame and, therefore, highly  blue-shifted by a factor $\sim \Gamma_{\rm bulk}^2$ \citep{Celotti07}. 
Given that  most  of these  photons have energies close to the Hydrogen  Ly$\alpha$ transition
and  assuming that the jet bulk Lorentz factor is of the order of  $\Gamma_{\rm jet}\sim$10,
the BC emission is expected to take the form of a variable thermal emission in the soft X-ray \citep{Sikora97, Celotti07}.

The expected transient character of the BC emission is due to the presence of local overdensities within the jet, 
which are launched very close to the central engine and accelerate up to hundreds or thousands  of  accretion radii  \citep{Celotti07, Kataoka08}. 
Since the high-energy flares observed in blazar jets are thought to be produced by the collision of shells flowing at different 
velocities, BC emission is expected as a precursor of $\gamma$-ray flares \citep{Moderski04}.

From an observational point of view, in spite of the fact that,  in this particular energy band,
FSRQ blazars usually  show  a minimum of the beamed radiation, BC emission have never been clearly and conclusively detected.
\cite{Celotti07} invoked BC emission to explain the departure from simple power law  of the X-ray spectrum of GBB1428+217 at z=4.72;
\cite{Kataoka08}, \cite{DeRosa08} and \cite{Kammoun17}  detected kT$\simeq$1 keV thermal components in 
the X-ray spectra of the blazars  PKS1510-089 (z=0.361), 4C04.42 (z=0.965) and  4C+25.05 (z=2.368) respectively.
However no clear evidence of the expected variability has been observed so far.
\begin{figure}
\center{
\includegraphics[width=7.5cm] {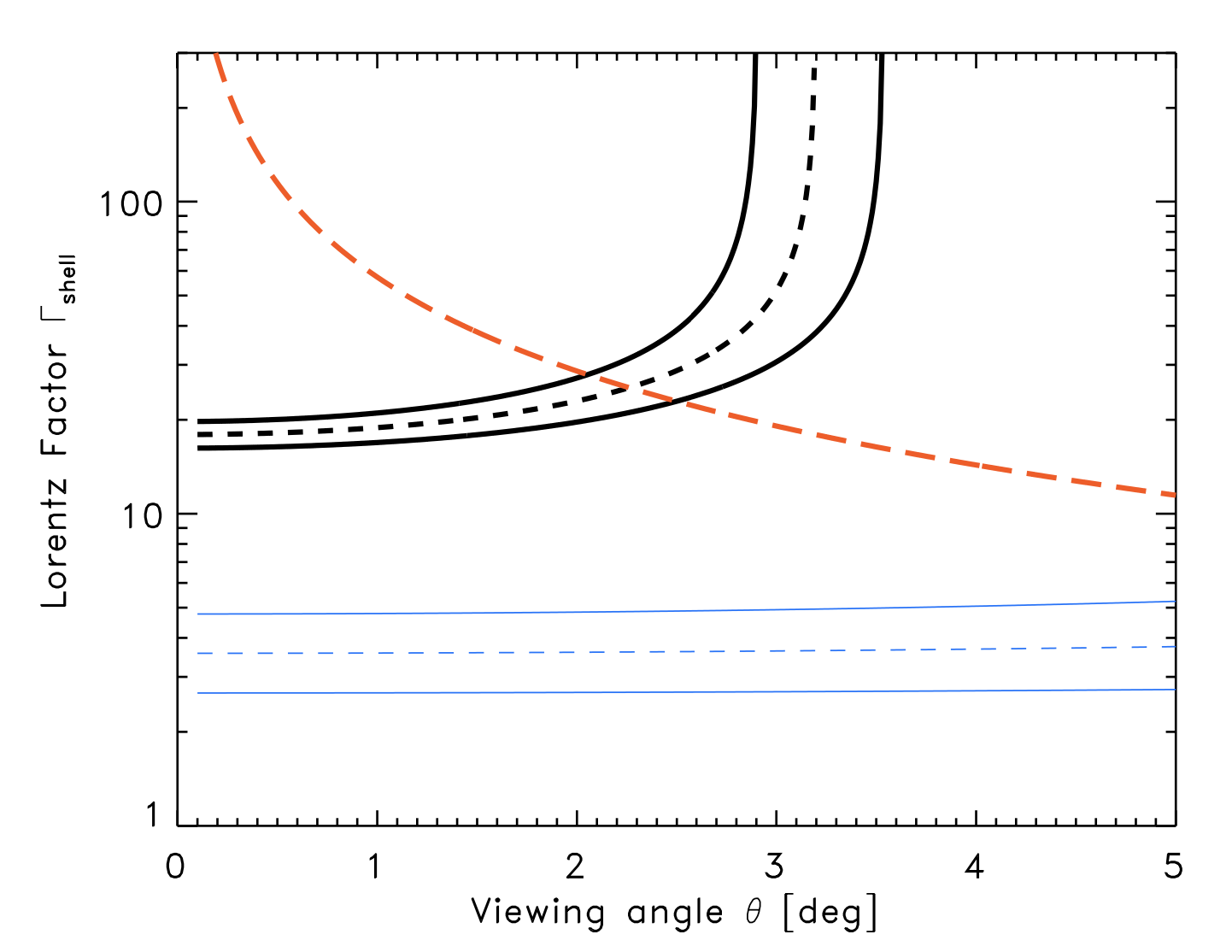}      
\caption{Black lines show the allowed values of the Lorentz factor $\rm \Gamma_{shell}$  as function of the  line of sight angle $\rm \theta_{view} $. 
Blue lines are the values allowed by the core-dominance argument for the $\rm \Gamma_{jet}$ in the VLBI observation \citep{Spingola20}. 
In both cases dashed lines represent the best values, with continuous lines being the 1$\sigma$ uncertainties. 
Red dashed line  shows the relation $\rm \Gamma = \frac{1}{\theta_{view}}$. 
} 
\label{fig:gammateta}
}
\end{figure}

\subsection{Bulk comptonization of broad line photons}
    
As said, the main contribution to BC emission is expected by  BLR photons \citep{Celotti07}. 
In this scenario the observed black-body  would have a mean energy 
\begin{equation} \label{eqn:bbkt}
\rm  KT_{\rm BLR,OBS} = \delta~\Gamma_{\rm shell} KT_{\rm Ly\alpha } ,
\end{equation}
    
where $\delta$ is the Doppler factor  at the viewing angle $\rm \theta_{view}$, ${\rm \delta = \frac{1} {\Gamma( 1-\beta~cos\theta_{view}) }  }$.
KT$_{\rm Ly\alpha }$ is the mean energy of black body peaking at a Ly$\alpha$ energy,
which is $ {\rm KT_{Ly\alpha }  = \frac {h\nu_{Ly\alpha }} {2.82}  }$ \citep{Celotti07}.
   
The expected observed luminosity is:
\begin{equation}
{\rm  L_{BC}=\frac{4}{3} c \sigma_{T} U_{BLR} \Gamma^{2} \delta^{4} N_{e}},
\end{equation}
   
where  $\rm{N_{e}}$ is the number of electron \citep{Moderski04} and 
    ${\rm U_{BLR}}$ is  the energy density of the BLR line emission.
This latter can be approximated by   
${ \rm U_{BLR} =  \frac{1}{12\pi} erg ~cm^{-3}},$ uniform in the region within  the BLR radius and
independent  of the disk luminosity and  BLR size \citep{Celotti07, Ghisellini13}. 
This yields 
\begin{equation}\label{eqn:lumi}
 {\rm  L_{BC}=7.08 \times 10^{-16} erg ~ s^{-1} \Gamma^{2} \delta^{4} N_{e}}.
\end{equation}

The present observation of PSO J0309+27 shows two soft flares, 10 ks apart, 
with similar characteristics. Given the observed KT$_{\rm BB}$=0.33$^{+0.07}_{-0.06}$,  on average, the two shells 
would have 
\begin{equation} \label{eqn:bbdata}
\rm \delta~\Gamma_{shell}  = \frac {  KT_{BB} \times (1+z) }  {KT_{Ly\alpha } } =  642_{-116}^{+136} .
\end{equation}
   
The region of the ${\rm \Gamma-\theta_{view}}$ plane satisfying this condition  is shown in  Fig.~\ref{fig:gammateta}.  
The constraint on the $\rm \delta~\Gamma_{shell} $ product sets an upper limit  on the viewing angle
$\rm \theta_{view} \lesssim$ 3$^{\circ}$. 
Approaching this limit  the Lorentz factor is constrained to be ${\rm \Gamma_{shell}}  \gtrsim $ 20.
With $\rm \theta_{view} < 2^{\circ}$, ${\rm \Gamma_{shell}} $ is limited to $\sim$ 20.  
\begin{figure}
\center{
\includegraphics[width=7.5cm] {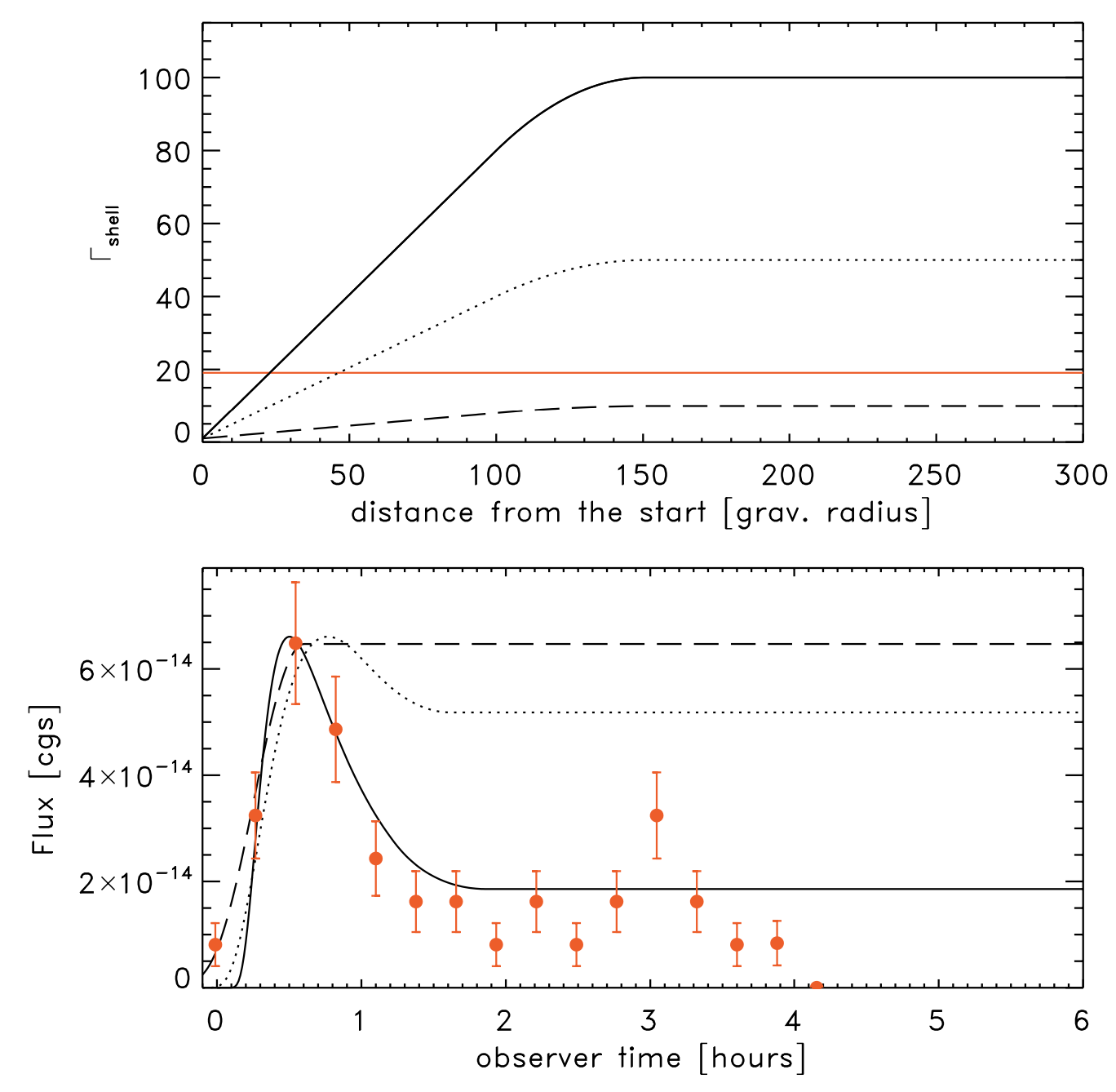}      
\caption{{\bf Upper panel:} Three different shell velocity profiles assumed to compare the prediction of the BLR photons BC 
model with the observation. In all the cases the acceleration is uniform up to a maximum value, and from that,
the shell is assumed to proceed at constant speed, with 
$\rm\Gamma_{shell}$=100, 50  and 10 respectively, reached in a space of $\sim$ 100 gravitational radii. 
The red line shows the value  $\rm \frac {1} { \theta_{view}} $, assuming $\rm \theta_{view}$  = 3.0$^{\circ}$.
{\bf Lower panel:} With the assumption of  $\rm \theta_{view}$ = 3.0$^{\circ}$ we show  
the BC light curve produced in the 0.5 -2.0 keV in the three cases  with the black continuous, dotted  and dashed lines respectively, 
compared to the observed data, shown in red. 
The light curves have been normalized to match the observed flux and  have a maximum when $\rm\Gamma_{shell} \sim  \frac {1} { \theta_{view}} $.
} 
\label{fig:bcblr}
}
\end{figure}

In the model presented by \cite{Celotti07}  the flux increase is caused by the shell acceleration, with the decline being 
due to the BLR energy density ($\rm U_{BLR}$) drop occurring at the limit of the BLR.
Unlike this picture, in our  data, the observed flare duration $\rm \Delta \tau_{obs}$ 
places very tight limits on the size of the  emitting region R$\rm_X$. 
Indeed causality  requires that 
$\rm R_{X} \lesssim c  \Delta\tau  $, where  
$\rm \Delta\tau = \Delta \tau_{obs} \delta~\Gamma_{shell}$ \citep{Ghisellini13}. 
Assuming the $ {\rm \delta~\Gamma_{shell}} $ value constrained by the observed BB energy, yields 
$\rm  R_X ~\sim 10^{16} cm$ equivalent to $\sim$ 100 gravitational radii,  a factor 20 smaller than the estimated  
size of the BLR of PSO J0309+27 (see Sect. 2).

The shape of the observed light curve might be due to the beaming effect of an accelerating shell.
In fact, the observed luminosity depends on the factor  $\Gamma^{2} \delta^{4}$ (Eq. \ref{eqn:lumi}).
At a given viewing angle $\rm \theta_{view} ( >0)$, the Doppler factor $\delta$ of an increasing  $\rm\Gamma_{shell} $  would peak
when $\rm \Gamma_{shell} \sim \frac {1} { \theta_{view}}$; at higher value of the Lorentz factor $\rm\Gamma_{shell} $ the observed emission drops 
with $\rm \theta_{view}$  being larger and larger than $\rm \frac {1} { \Gamma_{shell} }$ \citep{Ghisellini13}. 
   
In order to compare data and predictions (Fig.~\ref{fig:bcblr}), we assumed a $\rm \theta_{view}$=3$^{\circ}$, as required by the Eq.~\ref{eqn:bbdata}  condition
 when $\rm  \theta_{view} \gtrsim \frac {1} {\Gamma_{shell}}$ (Fig.~\ref{fig:gammateta}) and, for simplicity, a uniformly accelerating
shell, which is $\Gamma(x)\propto x$.
The predicted light curve  is calculated integrating in the (0.5-2.0) keV band  the BB spectrum with the mean energy varying according 
to the  Eq.~\ref{eqn:bbkt}.
We found that  a shell accelerating up to $\rm \Gamma_{shell} \sim$ 100 in a space equal to $\sim$ 100 gravitational radii may match the data, 
whereas a lower  $\rm \Gamma_{shell}$ maximum value would fail reproducing the descending slope  (Fig.~\ref{fig:bcblr}).
A normalisation consistent with the observed flux is provided by ${\rm N_{e} \simeq 5.0 \times 10^{53}}$.
If we make the assumption that the shell is loaded with the same amount of protons and electrons,
 this would mean a shell mass  M$_{\rm shell} \sim $8$\times 10^{29}$g flowing  along the jet 
 in a 160 ks time scale ($\rm \Delta\tau = \Delta \tau_{obs} \delta~\Gamma_{shell}$). 
 Given the estimated mass of 8$\times$10$^8$ solar masses, a luminosity of 0.3 times the Eddington limit  (Belladitta et. 2021, in preparation)
 and assuming a 10\% radiative efficiency, this rate would be equivalent to 10-15\% of the  SMBH accretion rate. 
This would mean that, in limited time intervals, a not negligible fraction of the matter usually accreting from the disk onto the SMBH is launched 
 along the jet. 

Because the reference model is highly uncertain and  we chose it arbitrarily, we limited ourselves to a rough and qualitative comparison, 
without trying to estimate the best fit parameters. 
 A physically motivated jet dynamical model, encompassing such an extreme condition is beyond the goal of the present
work. Here we just note that a similar shell dynamic has been hypothesized in the case of the FSRQ 3c279,
with mass similar to PSO J0309+27,  to explain a few hundred seconds flux variation in the $\gamma$-ray ($>$ 100 MeV) 
emission \citep{Ackermann16}.
   
The shell Lorentz factor value  ${\rm \Gamma_{shell}}$, estimated in this context, is significantly  higher  than
the upper limits set by the VLBI  data (Spingola et al. 2020) which constrain  the $\Gamma_{\rm jet}$ mean value, 
between 3-5 as long as  $\rm \theta_{obs} \lesssim$ 4$^{\circ}$ (Fig.~{\ref{fig:gammateta}}). 
With $\Gamma_{\rm jet}$ values at this level,  the BC emission would remain confined in the UV band; 
it becomes observable when faster shells come into action with Lorentz factor values ${\rm \Gamma_{shell}}$ high 
enough  to bring UV photons in the soft X-ray regime and producing spectral variations in observable timescales.
\begin{figure}
\center{
\includegraphics[width=9cm] {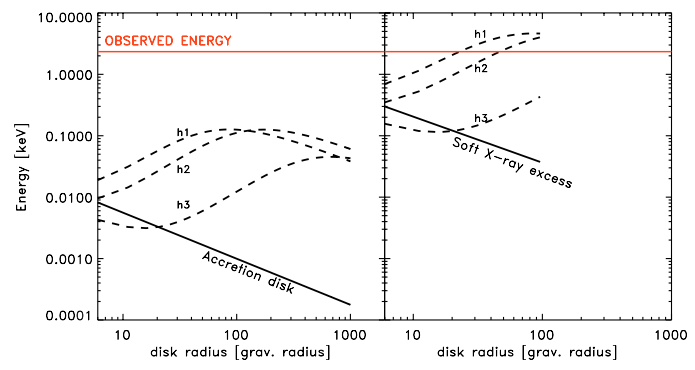}      
\caption{{\bf Left panel:} the expected  temperature profile of the accretion disk (continuous line) and 
the expected energies after BC at 3 positions in the jet (dashed lines; h1 = 50, h2 =100 , h3 =400 gravitational radii) , assuming $\Gamma$=20.
{\bf Right panel:} the same of the left panel, with a temperature profile peaking at 0.3 keV  and extending for only 100 gravitational radii in order 
to  naively depict the X-ray Soft excess.} 
\label{fig:bcdisk}
}
\end{figure}
\subsection{Bulk comptonization of disk photons}
   
Since the observed flux variation timescale requires that the emission region has to to be much smaller than the BLR radius, 
photons directly coming from the accretion disk (AD) might be suitable seeds for the bulk comptonization.
The disk emission is expected to be the superposition of  series of  black-bodies 
with mean energy dependent on the distance from the center according to  r$^{-0.75}$  \citep[][SS73]{Shakura73}.
Given the mass and the accretion rate of PSO J0309+27 (see Sect. 2), KT$\rm_{Disk}$(r) $\lesssim$ 10eV.
If involved in the BC process, disk photons should be observed with 
\begin{equation}
\rm  KT\rm_{obs}(r) = \frac {1-\beta~cos\theta_{disk}} {1- \beta~cos\theta_{view}}  KT_{Disk} (r) 
\end{equation}
where $\rm \theta_{disk}$ is the angle between the jet and the directions of the incident photons , with 
$\rm \theta_{disk}$ = 0 when photons and jet have same direction \citep{Celotti07}. 
At small $\rm \theta_{view}$ and with high velocities ($\beta \sim$1), as in the present case, the 
growth factor could reach values of some hundreds even for $\theta_{disk} < \pi$/2.  
On the other hand the highest energy photons come from the innermost part of the disk, 
where the  $\rm \theta_{disk}$ angle is smaller and the energy gain is lower. It follows that,  
cold jet electrons would not be able to up-scatter UV photons to the observed  X-ray energies (Fig.\ref{fig:bcdisk}),
not even with fast shell like those hypothesized in the previous section.
   
However, it has been observed that,  beside the SS73 disk, many QSOs are characterised by the presence of a hotter component, 
the so called X-ray soft excess, with typical mean energies in the 0.1-0.3 keV range \citep[e.g. the case of 3c273, ][]{Page04} 
and with luminosities equal to some fraction of the disk ones.
The origin of this component is not clear. A popular possibility is that it could be emitted from a comptonizing region lying on the
central part of the accretion disk, like a "warm skin" , \citep{Janiuk01, Done12}. 
The IC scattering of these photons by jet cold electrons  would explain, at least 
as a  first approximation, both the energy  and the time-scale of the observed transient emission. 
In this scenario, in addition to the shell velocity curve, the predictions strongly  depend on the geometry of the region 
responsible for the  X-ray soft excess. In Fig.\ref{fig:bcdisk} we draw a very basic scheme with the only aim of roughly estimating 
the energies of scattered photons.  In this picture, the shell Lorentz factor values, here required, are
significantly lower than the ones required by scenario in which the BC is due to the BLR photons.
    
It is reasonable to think that the same cold electrons responsible for the disk photons comptonization  
must subsequently  up-scatter the BLR photons as well. However, this would produce an increase of the soft X-ray flux 
on a timescale of $\sim$ 2 days, much longer than the our observation.

 \subsection{The case of QSO J074749+115352 }
    
Very recently, a few kilo-seconds timescale variation (3.8 ks) has been marginally detected in the
Chandra observation of the QSO J074749+115352 at z=5.26  \citep{Li21}.
Interestingly, this variability follows the same high-soft  /  low-hard  pattern we found for PSO J0309+27,
with very similar values of the spectral slopes.
If confirmed by statistically better data, the presence of a such a  variability in QSO J074749+115352, 
which is cataloged as radio-quiet QSO, would possibly undermine our interpretation as due to the jet BC emission.
However,  we note that  QSO J074749+115352 is clearly detectable in the VLASS survey (3.0 GHz) at  0.26'' distance 
from its optical position,  with a measured peak flux of  1.76$\pm$ 0.20 mJy  corresponding 
to a luminosity of $\sim$10$^{33}$ erg s$^{-1}$.
Since this level of luminosity is typical of radio-loud QSOs, we can conclude that, in the 
case of  QSO J074749+115352 too,  the observed emission is likely jet-contaminated.

\section{Summary and conclusions}
   
The two Chandra observations  performed 8 months apart point out a significant  
spectral variation of the emission of the blazar FSRQ PSO J0309+27. This change is entirely due 
to two soft flares present in the first observation with a similar observed duration of $\sim$ 250s,
rest frame.   
The extremely short time scale, together with the softer-when-brighter pattern make these events unique 
in the FSRQ observations reported so far in the literature.  
The bulk comptonization of UV external photons by the cold jet electron is a suitable  explanation.  
This process, although never observed,  is expected on a theoretical basis as a common feature in the emission 
produced by  leptonic jets.
We compared the Chandra data-set with the  model  predictions.
The observed energies  are  in agreement with the model which predicts that  
the major contribution to the BC emission is expected by the BLR photons. 
However, the observed time scale requires an emission region size much smaller than the BLR radius.  
We found that, in order to roughly reproduce our data,  the shell responsible for the
 BC emission, with BLR photons as seeds, should be accelerated up to $\Gamma_{shell} \sim 100$ 
 on a space scale of $\sim$10$^{16}$cm, equivalent to  $\sim$ of 100 gravitational radii.  

We also discussed the possibility that the BC seed photons come directly from the accretion disk.
This would be consistent with the observed data, if we assume the presence of a  soft X-ray excess, 
as observed in many low-redshift AGNs.      
   
Although not conclusive the present observations show that the study of the soft X-ray short time scale variability in the FSRQ 
emission might represent an excellent  tool to investigate not only the jet properties but also the innermost QSO environment,
which is opaque to higher energies. 
\begin{acknowledgements}

The scientific results reported in this article are based on observations made by the Chandra X-ray Observatory.
\end{acknowledgements}

\bibliography{tot20}{}
\bibliographystyle{aasjournal}



\end{document}